# An Open Model for Researching the Role of Culture in Online Self-Disclosure

Christine Bauer
Johannes Kepler University Linz
christine.bauer@jku.at

Katharina Sophie Schmid
University of Vienna
katharina.sophie.schmid@gmail.com

Christine Strauss
University of Vienna
christine.strauss@univie.ac.at

## Abstract

*The analysis of consumers' personal information (PI) is a significant source to learn about consumers. In online settings, many consumers disclose PI abundantly – this is particularly true for information provided on social network services. Still, people manage the privacy level they want to maintain by disclosing by disclosing PI accordingly. In addition, studies have shown that consumers' online self-disclosure (OSD) differs across cultures. Therefore, intelligent systems should consider cultural issues when collecting, processing, storing or protecting data from consumers. However, existing studies typically rely on a comparison of two cultures, providing valuable insights but not drawing a comprehensive picture.*

*We introduce an open research model for cultural OSD research, based on the privacy calculus theory. Our open research model incorporates six cultural dimensions, six predictors, and 24 structured propositions. It represents a comprehensive approach that provides a basis to explain possible cultural OSD phenomena in a systematic way.*

## 1. Introduction

Organizations are entering an era where real time data is available about their operations and their environments, offering new opportunities to increase their performance as well as enhanced chances to meet their customers' demands. Collecting data is no longer limited to an organization's internal processes or to its internal information flows: in fact, data is available about almost any aspect of its business. The key challenge for a successful enterprise is to transform information systems into intelligent systems that are able to manage the abundance of data and that are in accordance with the stakeholders' requirements and preferences.

A major source to learn about consumers is the personal information (PI) that they disclose about themselves. In fact, in electronic business-to-consumer relations, organizations typically require users to disclose PI, such as credentials for authentication, e.g. [34], contact and payment details for invoicing and payment for online purchases, e.g. [51], or information on the user's preferences for personalized recommender systems, e.g. [3], as well as personalized advertising, e.g. [9]. Moreover, many types of content that users generate on the Web (i.e. user-generated content) is PI that individuals disclose; for instance, textual postings and comments on online social networks or visual information such as photos or videos on respective platforms [18,46]. Social network services (SNSs) such as Facebook or YouTube would be nonexistent without having users disclosing PI [73] as providers of such services build their entire business on users' online self-disclosure (OSD). In short: OSD is a highly valuable source of information to sustain an organization's market position (e.g. for innovations, customization, marketing strategies, etc.). Consequently, the phenomenon of OSD has become an increasingly researched topic in various research threads such as information systems, e.g. [78], media psychology, e.g. [66], ethics, e.g. [26], or business economics, e.g. [31].

Still, it is not always favorable for users to provide PI openly. In fact, disclosing too much PI may have negative implications [2,4,39], including fraud, identity theft, violation of privacy rights, security attacks, or cyber-stalking [4,39,55,58]. As a consequence, people attempt "to manage the level of privacy that they wish to maintain" [67]. The so-called "privacy calculus theory" [23] has repeatedly been acknowledged as a suitable framework for studying OSD, e.g. [43]. This theory puts forth four determinants of OSD (i.e. *anticipation of benefits*, *privacy concerns*, *trusting beliefs*, and *risk beliefs*), which individuals weigh against each other and decide whether or not (and/or how much) to disclose [23].

The framework of "privacy as contextual integrity" [56,57] suggests that individuals adhere to norms that govern what is considered appropriate to reveal in which context. Since norms vary across groups, OSD behavior rendered appropriate or inappropriate is also likely to be differently manifested in norms.

HICSS

Fundamentally, the conception of privacy (which represents the substantial basis of the privacy calculus theory explaining OSD) varies from culture to culture [26,56]. Culturally-determined attitudes or beliefs may affect the user's decision whether or not to disclose PI.

However, most OSD studies either consider and analyze OSD on a worldwide uniform basis, e.g. [10], or study one-country samples without deeper cultural consideration, e.g. [40]. Still, several studies have shown that OSD differs across cultures, e.g. [11,41,62].

This culture-driven heterogeneity and its effects need to be analyzed in order to develop appropriate approaches to exploit but also to protect the users' PI. If disclosed PI differs across cultures, then the provided PI has different levels of breadth, depth, and validity. Furthermore, uniform interaction patterns with users will result in different degrees of OSD in various cultures. In this context, it seems important to note that organizations need to comply with privacy regulations (e.g. the Regulation (EU) 2016/679 of the European Parliament and of the Council of 27 April 2016 on the protection of natural persons with regard to the processing of personal data and on the free movement of such data [63]) when it comes to the exploitation of PI. In addition, corporate social responsibility implies responsible handling of PI [8,14]. Still, specific mechanisms and purposive activities may be necessary in initiatives for engaging in responsible and social actions to prevent individuals from disclosing "too much PI". In short, organizations need to adapt their strategies when using/exploiting PI from different countries.

In the context of OSD, Krasnova, et al. [43] provide significant insights into the role of culture; they investigate the effect of two out of Hofstede's six cultural dimensions [32,33]. Still, Bauer and Schiffinger [11] indicate that especially the prominently investigated individualism dimension is "quite far from being the most important culture-related moderator of OSD". They call for a research framework that allows for a bigger picture on the role of culture in the context of OSD.

Calling on this research gap, the present paper introduces an open research model for researching the role of culture in OSD, based on the privacy calculus theory [23] and the cultural dimensions by Hofstede [32,33]. We use the term "open research model" for the following reasons: *(i)* the model is open as we do not and cannot claim that the research model is exhaustive, as further influencing factors may be considered; *(ii)* it is a model and not a framework because we identify various impacts of predictor-moderator-effects; *(iii)* it is a research model and not a technical model because we formulate propositions (and do not describe a technical structure).

Research on the role of culture in online self-disclosure on a worldwide basis across various cultures is highly complex and requires a large set of extensive empirical studies. Accordingly, a consistent research model is significant to warrant homogeneity such that empirical results of different research teams with samples of different cultures will be valid and comparable. Our proposed open model provides such a basis.

The remainder of this paper is structured as follows: Section 2 provides the conceptual background of our research. We describe the concept of OSD, its predictors, and the impact of cultural dimensions on OSD. In Section 3, we develop 24 structured propositions based on findings from prior research. The literature basis builds on the ones provided in the meta-analyses by Bauer and Schiffinger [10,11] and was supplemented by targeted search for cultural aspects in OSD. The final section discusses and highlights our work's implications for further research and its practical use.

## 2. Background and Related Work

### 2.1. Online Self-Disclosure

Self-disclosure (SD) is defined as the communication of previously unknown personal, private information to others [17,37]. This may include facts about oneself, own experiences, or thoughts and feelings [17]. SD is an essential part of human communication (e.g. for forming and maintaining personal relationships), but also for validation of opinions, values, and perceptions [21,65]. It is generally regarded a risky act due to the intimacy of PI, which can lead to ridicule and rejection. This in turn can leave the discloser feeling awkward and vulnerable [60].

SD occurring online is referred to as online self-disclosure (OSD). Several studies have found that computer-mediated communication such as emailing and instant messaging exhibits different patterns of SD than, for instance, face-to-face (f2f) interactions. This is due to various factors such as anonymity, reduced non-verbal cues and more control over time and pace [48,72]. SD plays an important role in various fields of computer-mediated communication. It is used to establish legitimacy in SNSs, to reduce uncertainty experienced by others about oneself, or as a qualification for online services and transactions [28]. In an SNS context, SD refers to personal details, news, beliefs or even ideas shared on an online platform [43].

## 2.2. Predictors of Online Self-Disclosure

Various models have been used to explain OSD. Social exchange theory, for example, explains that individuals assess the costs and benefits of engaging in relationships. A relationship is considered worthy of engaging in, once perceived benefits outweigh associated costs [38]. Social penetration theory extends this view by taking into consideration the amount and nature of costs and benefits involved in SD. Among those are reciprocation (i.e. disclosing PI as a response to someone else's SD [7]) and vulnerability. Again, individuals analyze the risks and benefits involved and engage in SD if the assumed balance is positive for them [5].

Privacy calculus theory defines four determinants of a person's OSD: anticipation of benefits, privacy concerns, trusting beliefs, and risk beliefs. Its basic premise is that individuals assess the overall risk of engaging in OSD while taking anticipated benefits into consideration [23]. This model has been extensively used to examine OSD behavior in individuals [43].

Anticipated benefits refer to the rewards an individual expects to obtain as a result of disclosing PI. Among such rewards are enjoyment, social acceptance, and self-presentation [13,69]. Privacy concerns imply the fear of losing privacy after disclosing PI [79]. This can occur in an OSD context when another party is acting opportunistically. Trusting beliefs, on the other hand, relate to individuals trusting that their PI will be handled in a competent, reliable, responsible and safe manner. Lastly, risk beliefs refer to an individual's perception and awareness of opportunistic behavior of others, that might cause negative effects for the individual [23].

Additional factors derived from social exchange theory and social penetration theory affecting OSD such as perceived anonymity and perceived reciprocity have gained substantial attention in research as well [62]. Therefore, they will also be included into our research model; as a result, we incorporate six predictors in our open model, i.e. anticipated benefits, trusting beliefs, privacy concerns, risk beliefs, perceived anonymity, and perceived reciprocity.

## 2.3. Online Self-Disclosure in Different Cultural Contexts

The construct "contextual integrity" [56,57] ties protection for privacy to norms of specific contexts. What counts as private and what is considered appropriate to be revealed in a certain context may vary across cultures, as norms vary across cultures [56].

While OSD is mainly researched in a worldwide context (e.g. by studying a sample of Facebook users), it has also been subject to studies in a particular cultural context; for instance, investigating a Russian sample, e.g. [40] or one from Germany, e.g. [68].

Other studies compare two countries. For example, a survey interviewing Moroccan and US-American Facebook users revealed that Moroccans generally disclose less PI than Americans, as they perceive the damage incurred in case of violation as higher. Americans, in turn, showed lower privacy concerns than Moroccans [77]. Similar results were found regarding perceptions and behavior of Americans [41]. Overall, Americans were more involved in Facebook, felt more in control over sharing PI, and had greater trust in the SNS than Germans, who had less trust in the provider and felt less in control of how their PI was being handled. Another study using American and German participants also found that trust was a major determinant of SD decisions of Americans, while German participants based their decisions on privacy concerns [42]. By contrast, a survey conducted both f2f and online with American and Chinese participants showed that Americans anticipated more SD in f2f interactions than in online settings. A major source of concern was for Americans to have other members of their online community find out about their PI, whereas the Chinese participants were more concerned about third-party access to their PI from f2f communication [80]. In addition, some differences regarding user goals, self-expression, and interaction behavior online between different cultures were revealed [16].

However, studies dealing with SNSs and the role of culture have mostly focused on industrialized countries, e.g. [42,80]. Additionally, studies examining OSD have largely utilized small samples from certain segments such as university students, e.g. [25,27]. Moreover, the data used by these studies was mostly based on self-reporting instead of being behavioral in nature [40]. However, the scarce existing research has found that online interaction is indeed not culturally neutral [11,25,61,68].

## 2.4. The Impact of Culture on Online Self-Disclosure

The concept of culture is a very broad one, which is associated with local values, beliefs and traditions [32]. The arguably most frequently cited and widely accepted typology to differentiate between cultures has been presented by Hofstede [32,33]. His framework is based on data gathered from 116,000 IBM employees from over 70 different countries, which he collected between 1967 and 1973 [32]. Based on his findings, he proposed five distinctive dimensions to describe a country's culture, i.e. individualism/collectivism, power distance, uncertainty avoidance, masculinity/femininity, and long-term orientation. A sixth one, namely indulgence, was added later [33]. The scores of each country are not to be considered absolute values but rather as ones relative to other countries' scores. Additionally, they do not describe each individual's characteristics in their respective society but rather collective trends and tendencies [33]. Hofstede's dimensions have been found to not just be applicable in the offline world, but also in online interactions [45,75].

A few studies have explored the effect of culture on OSD and related aspects such as privacy concerns taken from privacy calculus [12,53]. For example, [43] have picked two dimensions from Hofstede's approach (i.e. *individualism* and *uncertainty avoidance*) and compared the behaviour of SNS users from the United States and Germany. Moderating effects of these two dimensions on the relation of the privacy calculus concepts and OSD have been analysed in their study and have been identified as pivotal determinants for further research on cultural differentiation, e.g. [30,43]; other dimensions are yet to be explored in research on the influence of culture in the context of OSD (see also [11]).

Overall, we claim that there is a need for a comprehensive research model that supports overcoming the complexity of the cultural impact on OSD. Only a consistent research model may warrant that studies with samples of different cultures, investigated by different research teams, will be valid and comparable. Thus, we substantiate a set of propositions based on extant theories and research in the following Section **Error! Reference source not found.** and propose an open research model.

# 3. Propositions for Research on the Role of Culture in Online Self-Disclosure

This section outlines the propositions (P1-P12c) concerning the predictors of OSD (Section 3.1) and the moderating influence of culture (Section 3.2). The open research model is visualized in Figure 1.

## 3.1. Predictors of Online Self-disclosure

Online users engaging in SD expect certain benefits from their activities, such as enjoyment or social acceptance [64,69]. Self-presentation and relationship maintenance are central benefits and drivers of OSD [23,41]. A positive balance of benefits and drawbacks from engaging in online activity favors OSD [23,38].

P1: *Anticipated benefits* have a positive effect on OSD.

Trust as another important factor in OSD is considered as a precondition for disclosing information [81]. Social exchange theory states that SD is more likely to occur if the relational partner is considered trustworthy [44,71]. This correlation has also been found for trust in online community members [62].

In e-commerce, consumers will most of the time only disclose themselves and thus make a purchase if they perceive a merchant as trustworthy [6]. Moreover, trust plays an important role in online communities where platforms are considered as reliable and trustworthy if PI is treated accordingly [62]. Trust may even out the negative impact of risk beliefs so that trusting parties may engage in risky behavior if a certain level of trust exceeds the level of perceived risk [23,29].

P2: *Trusting beliefs* have a positive effect on OSD.

Users feel more comfortable disclosing PI if the platform establishes higher levels of privacy among them [76]. Additionally, a study showed that privacy processes differ based on whether they are dispositional or situational [35]. The study also showed that the general privacy disposition has no effect on situation-related interpretation of trustworthiness, and that trust has a moderating effect on perceived privacy. Privacy concerns have also been named as reasons for refusing online transactions [19].

P3: *Privacy concerns* have a negative effect on OSD.
P4: *Risk beliefs* have a negative effect on OSD.

Studies have shown that the lack of personal identification decreases inhibition, which in turn leads to people sharing more PI than otherwise [20,59]. Also, people tend to disclose more PI in an online environment, as condemnation and rejection may not be attributed to them personally [49]. Thus, anonymity may lead to disinhibition among online users, which in turn makes them more likely to disclose PI.

P5: *Perceived anonymity* has a positive effect on OSD.

A key aspect of SD is reciprocity. Research has shown that people are much more likely to disclose PI if their communication partner disclosed PI about himself or herself earlier on. Such findings are backed by social penetration theory, according to which a person engaging in social interactions will share more PI in order to maximize perceived benefits [38]. This is also applicable for online interactions such as instant messaging [54]. A study by Posey, et al. [62] revealed that reciprocity had the greatest influence on OSD.

P6: *Perceived reciprocity* has a positive effect on OSD.

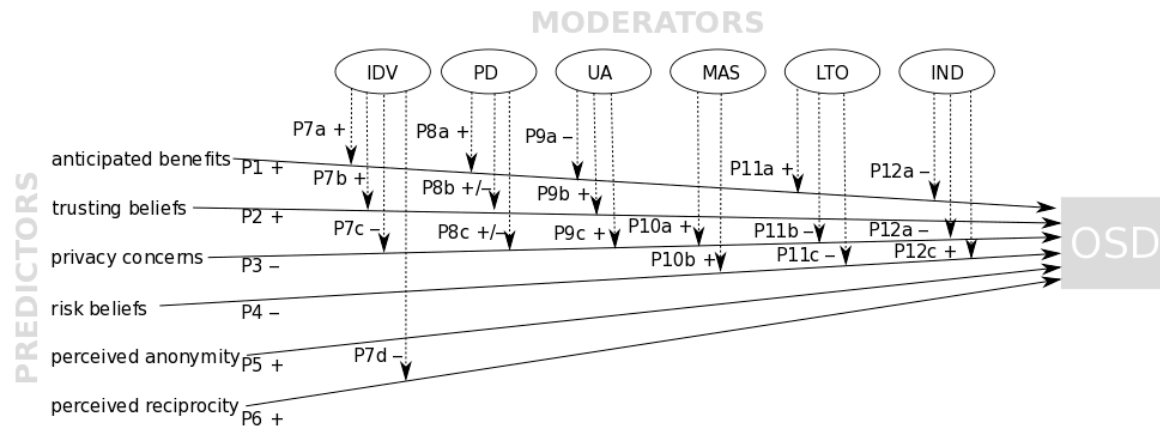

**Figure 1. An open model for researching the role of culture in OSD**

## 3.2. The Moderating Influence of Culture

*Individualism* (IDV) refers to a person's independence from collectivity and organizations, and looser ties between them. Collectivism stands for the opposite and implies stronger integration of individuals into groups and organizations [33].

As a consequence, certain traits such as hedonism and pleasure-seeking are attributed to individualistic cultures, where people prioritize their personal needs [22,74]. In an online context, it can thus be concluded that the effect of *anticipated benefits on SD* from engaging in online activities will be enhanced for cultures with higher levels of individualism due to their hedonic traits.

*P7a: Individualism* increases the positive effect of *anticipated benefits* on OSD.

The trust-formation process differs substantially between individualistic and collectivistic cultures. Collectivists focus on the predictability of future actions taken by the trustee, as well as his benevolence and the transferability of trust within a group. Individualists, on the other hand, calculate the costs and benefits of their interaction with a trustee [24]. Krasnova and Veltri [42] highlighted these assumptions by revealing that highly individualistic Americans put substantial emphasis on trust in SNS providers when making online decisions.

*P7b: Individualism* increases the positive effect of *trusting beliefs* on OSD.

While individualists attribute greater value to privacy, collectivists do not mind intrusion into their private life as much. Research has found that countries with higher levels of individualism also tend to be more concerned about their online privacy [47,53]. For example, Dinev, et al. [22] revealed that privacy concerns had greater influence on the use of e-commerce in more individualistic cultures. Paradoxically, it has also been reported that individualists on average share more photos online than collectivists [33]. Based on the abovementioned findings, the following can be stated about the moderating effect of *individualism*:

*P7c: Individualism* decreases the negative effect of *privacy concerns* on OSD.

Furthermore, it has been observed that collectivists are more likely to reciprocate to others than individualists [62]. This could be explained by the fact that collectivists put a stronger emphasis on social interaction in groups than to their personal independence from others [33].

*P7d: Individualism* decreases the effect of *perceived reciprocity* on OSD.

*Power distance* (PD) describes the acceptance of inequality of power in a country. Countries with greater power distance are thus more accepting of power inequality than those with a lower score. However, higher levels of PD have been found to be associated with greater mistrust and privacy concern [12,52,53]. Contrary to this, Cao and Everard [15] argued that countries with high levels of PD are less concerned about their privacy since they are used to authorities accessing their PI. Bauer and Schiffinger [11] found that PD increases the effect of *anticipated benefits* on OSD.

*P8a: Power distance* increases the positive effect of *anticipated benefits* on OSD.

P8b: *Power distance* moderates the effect of *privacy concerns* on OSD.

P8c: *Power distance* moderates the effect of *trusting beliefs* on OSD.

*Uncertainty avoidance* (UA) refers to a culture's attitude towards ambiguous and risky situations and whether it tries to avoid them. High levels thereof consequently imply greater concern, anxiety and stress. According to Bauer and Schiffiger [11], UA is among the two most influential cultural dimensions with respect to OSD (the other one being *indulgence*).

Our first proposition concerning UA (P9a) relates to its mitigating effect of *anticipated benefits* on OSD, which appears intuitive by nature and was also proven in earlier research [11].

P9a: *Uncertainty avoidance* decreases the positive effect of *anticipated benefits* on OSD.

Lim, et al. [45] posit that trust is strongly impacted by levels of both IDV and UA. The latter is assumed to be affected by pessimistic attitudes towards companies' incentives. Cultures with lower levels of UA tend to be less concerned about their privacy and are thus more likely to take risky actions on the basis of trust [24].

P9b: *Uncertainty avoidance* increases the positive effect of *trusting beliefs*.

Also, several studies have pointed out that higher levels of UA tend to lead to greater *privacy concerns* [12,52,53]. These findings emphasize the general concern that lies in the nature of cultures with high levels of UA.

P9c: *Uncertainty avoidance* increases the effect of *privacy concerns* on OSD.

The *masculinity/femininity* dimension (MAS) deals with gender roles and refers to whether a culture is more masculine or feminine. *Masculinity* is associated with men's assertiveness, materialism, success, less concern about others, and a stronger contrast to women's more gentle characteristics [32]. *Femininity*, on the other side of the continuum, describes cultures where both men and women tend to be rather tender, modest and concerned with quality of life [33].

MAS has also been examined in the context of *privacy concerns*: Milberg, et al. [53] have found a positive link between *masculinity* and *privacy concerns*. They concluded that stronger competitiveness leads to greater alert about misuse of PI. Krasnova and Veltri [42] came to a similar conclusion. Taking these findings into consideration, the following assumptions can be made:

P10a: *Masculinity* increases the negative effect of *privacy concerns* on OSD.

P10b: *Masculinity* increases the negative effect of *risk beliefs* on OSD.

The abovementioned reasoning by Krasnova and Veltri [42] and the underlying argument by Acquisti [1] concluding that striving for immediate benefits leads to higher OSD can also be used for explaining the effect of *long-term orientation* (LTO) on OSD: LTO might reinforce the effect of *anticipated benefits* on OSD. Although such an effect could not be shown in the study by Bauer and Schiffinger [11], we still postulate P11a, assuming a reinforcement of the positive effect. Furthermore, for the same reasoning by Krasnova and Veltri [42] and Acquisti [1], we expect that LTO augments the negative effects of privacy concern and of risk assessment on OSD; both impacts were shown in the study of Bauer and Schiffinger [11].

P11a: *Long-term orientation* decreases the positive effect of *anticipated benefits* on OSD.

P11b: *Long-term orientation* increases the negative effect of *privacy concern* on OSD.

P11c: *Long-term orientation* increases the negative effect of *risk beliefs* on OSD.

Bauer and Schiffinger [11] found in their analysis several moderating influences of *indulgence* on the privacy calculus theory predictors' effects on OSD. Their rather speculative line of argument states that *indulgence* also represents "control over one's life" [70], which explains their findings that *indulgence* reduces the positive effect of *anticipated benefits* (P12a), and tht it intensifies the negative effect of *privacy concerns* (P12b) and *risk beliefs* (P12c). Further research is necessary; for this reason, we include it in our open model of cultural OSD research.

P12a: *Indulgence* decreases the positive effect of *anticipated benefits* on OSD.

P12b: *Indulgence* increases the negative effect of *privacy concerns* on OSD.

P12c: *Indulgence* increases the negative effect of *risk beliefs* on OSD.

Figure 1 provides a synopsis of the above-described impacts of the predictors on OSD, together with the moderating effects of cultural dimensions. Positive and negative influences are visualized by arrows, which are tagged with the number of the respective proposition.

As shown in Figure 1, the relation of *privacy concerns* and OSD is influenced by the entire set of cultural dimensions, and the predictor *anticipated benefits* is also influenced by most of the moderators (except MAS). In the model, the relation between *perceived anonymity* and OSD is not influenced by any of the cultural attributes. Note that the relations depicted in Figure 1 are built on existing theories and/or results of empirical studies. Therefore, it shall not be concluded that there is no impact between a certain moderator and the relation between a certain predictor and OSD; it can only be concluded that (so far) there is no evidence for such an effect.

## 4. Discussion and Conclusion

PI disclosed by consumers online is a highly valuable source to learn about them. Not to mention, many online platforms would not even exist without the content provided by their users (i.e. user-generated content), including posts, comments, photos, and/or videos. As providers of such services largely rely on users' OSD, it is crucial for such organizations to study and understand users' OSD. For users, yet, it is not always favorable to provide their PI openly due to, among others, privacy and security reasons.

While organizations tend to consider a "culturally universal Internet user", we emphasize that, with respect to OSD, cultural uniformity does not exist. Consequently, organizations need to address their stakeholders differently across countries to be capable of handling their PI accordingly. This implies separate analysis of PI data across various countries and requires organizations to interact with users differently as users' OSD is affected by their culture.

Our open research model represents a comprehensive approach that provides a basis to perform cultural OSD research and to explain possible cultural OSD phenomena in a systematic way. It is built on several sub-approaches, which we have consolidated. These sub-approaches have proven to be reasonable in well-defined settings; furthermore, it incorporates novel aspects together with new cause-and-effect chains largely underpinned by recent research. Our suggested open research model is propositions-driven and incorporates six moderators that affect the relation of six predictors on OSD. Thus, we suggest and substantiate 24 structured propositions. As it is an open model, it can be further developed and extended in continuous research endeavors.

The suggested propositions may be examined through laboratory experiments, survey studies, and/or field experiments. The sample population and their cultural background deserve careful choice and design so that conclusions can be generalized.

Further research on the cultural role in OSD may be performed in three major directions, i.e. method, provision of evidence, and applications.

Future method-oriented work may be twofold: focusing on *(i)* the predictors and *(ii)* the moderators. Our model is based on the assumption that the suggested predictors are independent from each other and that they have no mutual effects or interdependencies; this needs to be substantiated by further empirical research. Another predictor-relevant research question refers to the completeness of the number of suggested predictors. Additional predictors should be incorporated into the model if further empirical studies introduce them; their independence of every other predictor in the model should be examined.

Further method-oriented work refers to the moderators: we have chosen Hofstede's work on cultural attributes because his studies provide up-to-date numerical evidence that may be highly suitable to make transparent cultural distinctions in a certain context. Hofstede's approach has been criticized regarding conception, methodology, and interpretation [36,50]; for instance, for equating *culture* with *nation* and for disregarding ethnic aspects. Still, this approach seems to be appropriate for the purpose of business applications as companies usually target markets on a national or supranational level. Applying any other moderators' scheme is generally possible, but implies a re-evaluation of impacts (i.e. what effects are to be expected on which predictor-OSD-relation). With respect to the framework of contextual integrity that allows for defining norms on various group levels, this seems a promising perspective that we intend to pick up.

The second major direction of further work refers to the provision of evidence through empirical studies. Rather than performing research on a single country or making pairwise comparisons using one or two moderators, broader studies (in terms of cultures and/or of Hofstede's attributes) will lead to better quality in results and are to be preferred.

The third major direction of further work refers to applications, as our model has several benefits for the industry: for instance, companies that intend to launch marketing campaigns or products that are based on OSD or involve OSD (e.g. SNS applications) in new markets may perform an analysis to get well-founded insights into specific cultural attitudes of future customers. Future work should come up with frameworks and guidelines to support organizations in their international business activities.

Since concerns act as a barrier to protect users from disclosing too much PI and since users' tendencies to disclose PI are further tied to their cultural background, organizations need to include varying measures and actions depending on their users' native countries. Similarly, PI disclosed by users from different countries will require different measures for and degrees of data protection, since, depending on the cultural background, a user is more or less inclined to disclose PI in detail.

## 5. Acknowledgments

This research is supported by the Austrian Science Fund (FWF): V579.